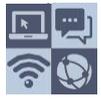

*future internet*

MDPI

*Article*

# Internet of Nano-Things, Things and Everything: Future Growth Trends

**Mahdi H. Miraz** [1] [ID], **Maaruf Ali** [2], **Peter S. Excell** [3,*] **and Richard Picking** [3]

1   Centre for Financial Regulation and Economic Development (CFRED), The Chinese University of Hong Kong, Sha Tin, Hong Kong SAR, China; m.miraz@cuhk.edu.hk
2   International Association of Educators and Researchers (IAER), Kemp House, 160 City Road, London, EC1V 2NX, UK; maaruf@ieee.org
3   Faculty of Art, Science and Technology, Wrexham Glyndŵr University, Wrexham LL11 2AW, UK; r.picking@glyndwr.ac.uk
*   Correspondence: p.excell@glyndwr.ac.uk; Tel.: +44-797-480-6644

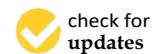

**Abstract:** The current statuses and future promises of the Internet of Things (IoT), Internet of Everything (IoE) and Internet of Nano-Things (IoNT) are extensively reviewed and a summarized survey is presented. The analysis clearly distinguishes between IoT and IoE, which are wrongly considered to be the same by many commentators. After evaluating the current trends of advancement in the fields of IoT, IoE and IoNT, this paper identifies the 21 most significant current and future challenges as well as scenarios for the possible future expansion of their applications. Despite possible negative aspects of these developments, there are grounds for general optimism about the coming technologies. Certainly, many tedious tasks can be taken over by IoT devices. However, the dangers of criminal and other nefarious activities, plus those of hardware and software errors, pose major challenges that are a priority for further research. Major specific priority issues for research are identified.

## 1. Introduction

The applications and usage of the Internet are multifaceted and expanding daily. The Internet of Things (IoT), Internet of Everything (IoE) and Internet of Nano-Things (IoNT) are new approaches for incorporating the Internet into the generality of personal, professional and societal life, plus the impersonal world of inanimate quasi-intelligent devices. This paper examines the current state of these technologies and their multidimensional applications by surveying the relevant literature. The paper also evaluates the various possible future applications of these technologies and foresees further developments and how these will both challenge and change the way that future life will be lived. This paper presents an update on our previous work [1] presented at the Internet Technologies and Applications Conference in 2015 (Wrexham, UK) by extending the survey duration to reflect the current technological advances since 2015. New dimensions of discussion have also been added such as the future challenges IoT is currently facing. The discussion on IoT, in Section 2, has been further expanded by adding sub-categories of IoT based on the scope of its usage as well as the components of typical IoT systems, with a listing of the top ten IoT segments for 2018 based on a survey of 1600 enterprise IoT projects. The discussion on IoNT has been augmented by the inclusion of discussion of the Internet of Bio-Nano-Things (IoBNT), limitations and challenges of IoNT and presentation of examples of earlier research advances in the field. The deliberation on "Future Internet" has been extended as well as updated to reflect new research, associated challenges and future trends. Section 6, namely "Challenges and Impediments to IoT", has been added, scrutinizing 21 of the most significant current and future challenges.

The paper first provides a critical discussion on IoT, IoE and IoNT in Sections 2–4 respectively. Section 5 portrays the Future Internet that is predicted to be mediated by adoption of IoT. Challenges and Impediments to IoT are covered in Section 6. Section 7 ends the paper with up-to-date concluding discussions.

## 2. Internet of Things (IoT)

The term "Internet of Things" or "Internet of Objects" has come to represent electrical or electronic devices, of varying sizes and capabilities, that are connected to the Internet, but excluding those primarily involved in communications with human beings, i.e., the traditional Internet. The scope of the connections is ever broadening beyond basic machine-to-machine communication (M2M) [2].

IoT devices employ a broad array of networking protocols, applications and network domains [3]. The rising preponderance of IoT technology is facilitated by physical objects being linked to the Internet by various types of short-range wireless technologies such as: RFID, UWB, ZigBee, sensor networks and through location-based technologies [4]. The emergence of IoT as a distinctive entity was achieved, according to the Internet Business Solutions Group (IBSG), actually when more inanimate objects were connected to the Internet than human users [5]. According to this definition, this occurred in mid-2008. This is an accelerating ongoing process, especially with the rollout of Cisco's "Planetary Skin", the Smart Grid and intelligent vehicles [5]. IoT will make the impact of the Internet even more pervasive, personal and intimate in the daily lives of people.

IoT devices are not currently strongly standardized in how they are connected to the Internet, apart from their networking protocols; however, this could be a relatively short-term inhibiting factor. IoT may be employed with added management and security features to link, for example, vehicle electronics, home environmental management systems, telephone networks and control of domestic utility services. The expanding scope of IoT and how it can be used to interconnect various disparate networks is shown in Figure 1 [5].

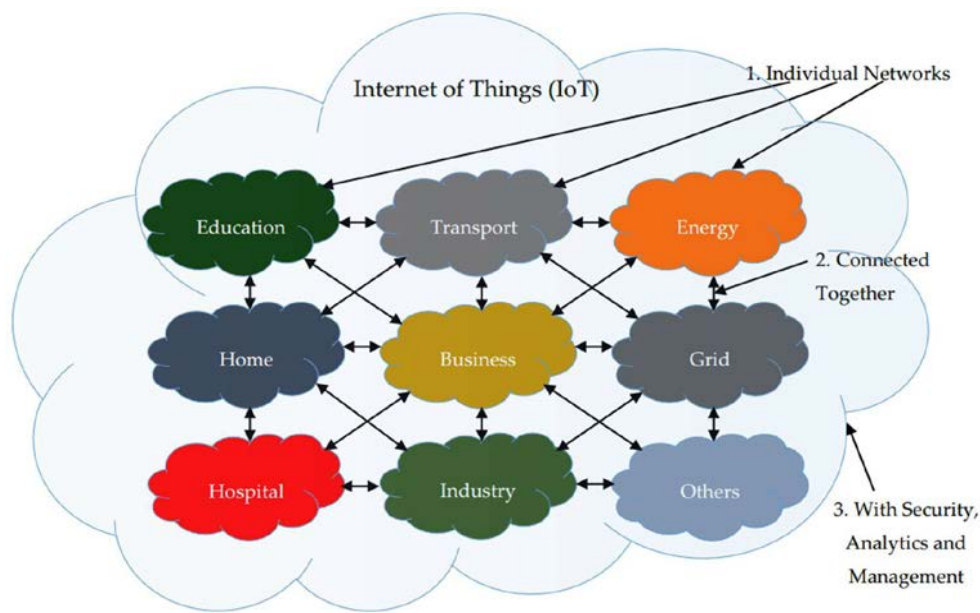

**Figure 1.** IoT can be viewed as a Network of Networks. Adapted from [5].

Based on the type of use, IoT can be further categorized as Industrial Internet of Things (IIoT) and Consumer Internet of Things (CIoT), alternatively known as Human Internet of Things (HIoT), as shown in Figure 2:

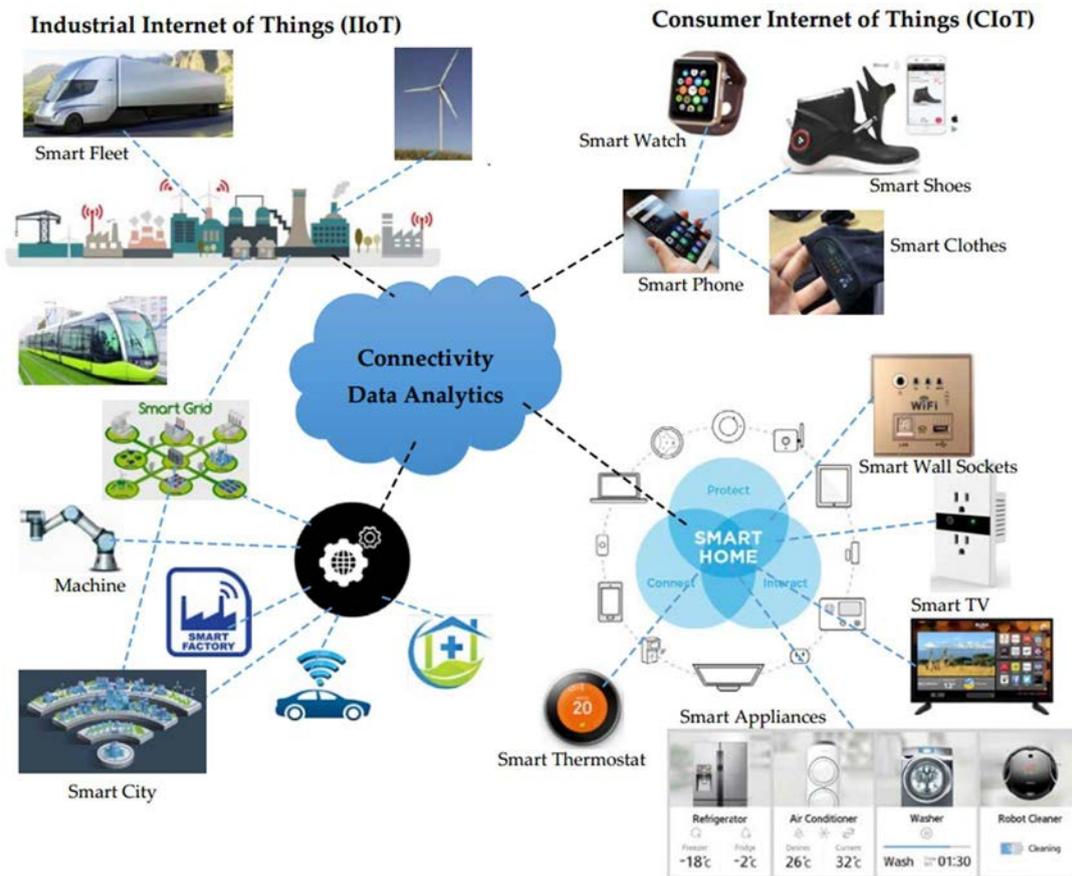

**Figure 2.** IIoT and CIoT, adapted from Moor Insights & Strategy's Report [6].

A generic IoT system typically consists of five components, which are:

(1) Sensors: which are used to mainly collect and transduce the data;
(2) Computing Node: a processor for the data and information, received from a sensor;
(3) Receiver: to facilitate collecting the message sent by the computing nodes or other associated devices;
(4) Actuator: based on the decision taken by the Computing Node, processing the information received from the sensor and/or from the Internet, then triggering the associated device to perform a function;
(5) Device: to perform the desired task as and when triggered.

As an example, Figure 3 lists the 2018 top ten IoT segments, compiled by Scully [7], who mined the Web to identify 1600 actual enterprise IoT projects, based on a strict definition.

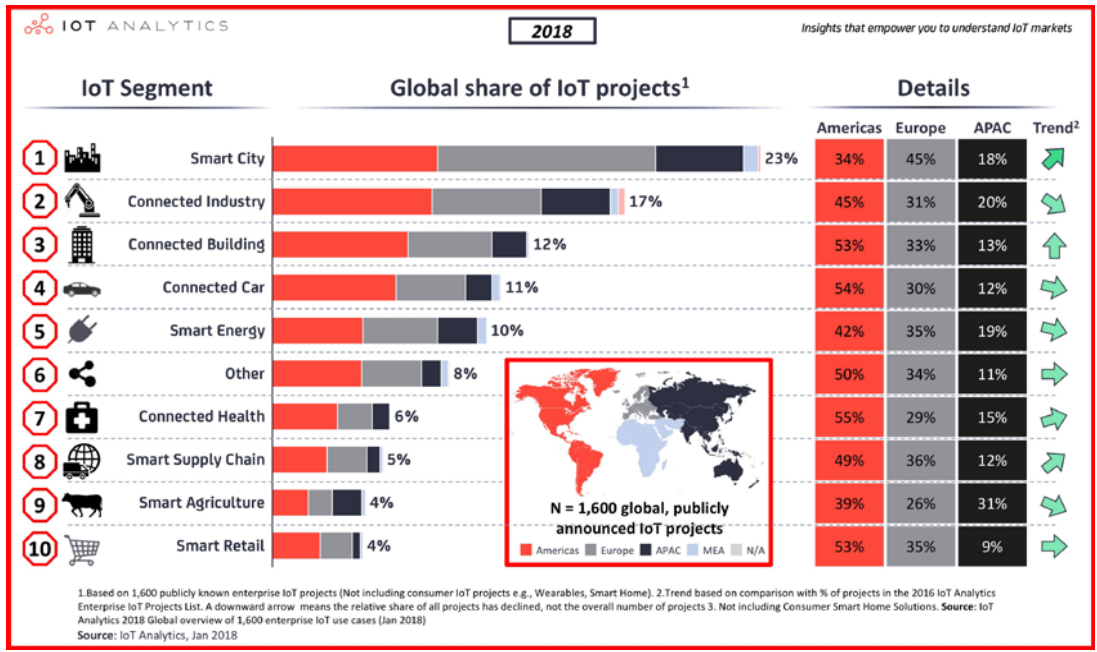

**Figure 3.** The top ten IoT Segments in 2018—based on 1600 real IoT projects [7].

## 3. Internet of Everything (IoE)

Both Cisco and Qualcomm have been using the term IoE [8,9]. However, Qualcomm's interpretation of the term has been replaced by the IoT by a majority of others. Cisco's usage has a more comprehensive meaning. The Cisco version of IoE is built upon the "four pillars" of people, data, process and things, whereas IoT is only composed of "things", as shown in Figure 4. IoE also extends business and industrial processes to enrich the lives of people.

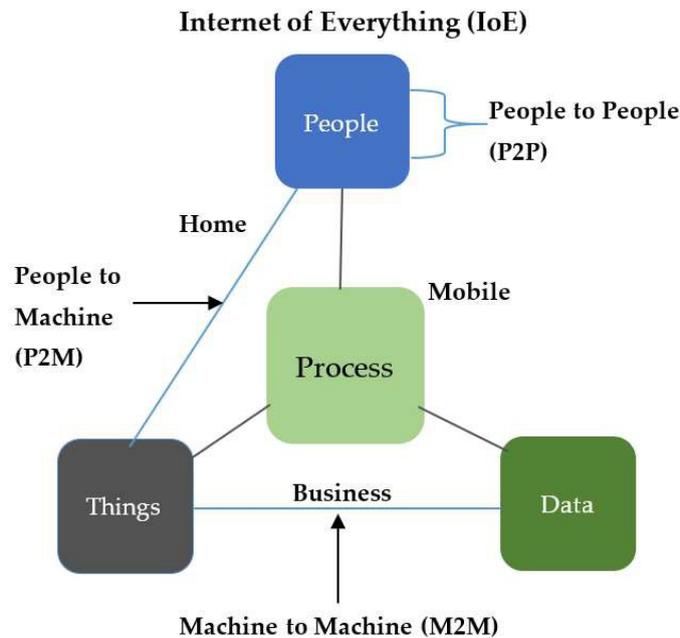

**Figure 4.** Internet of Everything (IoE). Adapted from [10].

The independent, non-networked and unconnected devices of the past are now being connected to the Internet, including machine-to-machine (M2M), person-to-machine (P2M), and person-to-person (P2P) systems. This enveloping of people, processes, data and things by IoE is shown in Figure 4 [9,10].

The Futurist, Dave Evans, states that, rather than simply "things", the issue is more about the "connections among people, process, data, and things" that is at the heart of the Internet of Everything and creates the "value" [11]. Qualcomm's CEO, Steve Mollenkopf, stated in 2014 that IoT and IoE were "the same thing" [8].

According to Cisco, many organizations are going through growth waves of S-curves, as shown in Figure 5. These IoT growth waves are leading to the eventual actualization of the complete IoE [9,12]. With each successive wave of added features and greater network connectedness—this leads to expansion of the IoE, with its many novel opportunities as well as risks [13]. Interestingly, this interpretation of progression through a succession of S-curves correlates closely with the model for accelerating change proposed by Raymond Kurzweil, which is also based on successive S-curves [14].

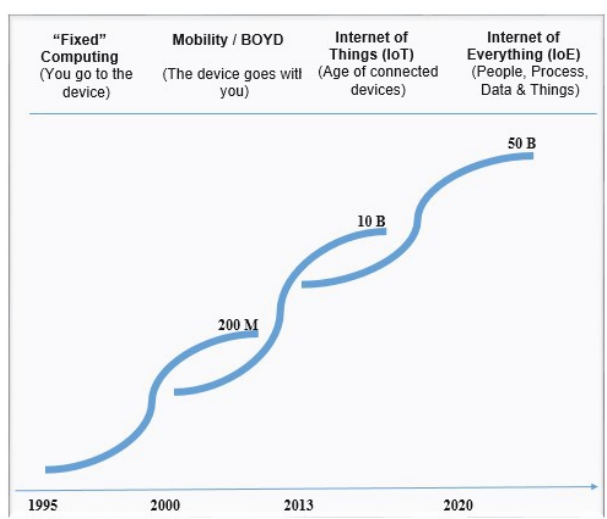

**Figure 5.** Internet growth is occurring in S-curve waves [9,12].

The IoE has the potential to extract and analyze real-time data from the millions of sensors connected to it and then to apply it to aid "automated and people-based processes" [15]. Other benefits include the use of IoE in helping to achieve public policy goals, environmental sustainability, economic and social goals [15].

Traditional office-based applications such as financial trading have now moved into the domain of the mobile platform with the use of smartphones, as well as many other applications, aided by IoE [16,17]. The application of IoE is facilitated by the expansion of Cloud Computing, helping to connect "everything" online [18]. A study by Cisco in February 2013 predicted that $14.4 trillion may be exploited in the next ten years by implementing IoE with M2M, M2P and P2P [18].

Cities, which in the future may be regarded as a scaled version of the IoE, will benefit the most from being connected in terms of using information intelligence to address city-specific concerns [19]. This will become more so as cities become "Smart Cities" [19], utilizing IoE together with "Big Data" processing [20]. Examples include monitoring the "health" of highways and attending to their repairs using road-embedded sensors, road traffic flow control, agricultural growth monitoring, education and healthcare [21,22]. The future is most likely to see cities become "Smart + Connected Communities", formed using public-private partnerships to help enhance the living conditions of the citizens.

As urbanization continues to increase, predicted to be 70% by the 2050s [21], the use of IoE will become almost critical in implementing such features of the future city as the Smart Grid and automation of traffic planning and control [19]. IoE is also forming a foundation in the mining industry of fossil fuels and in remote monitoring, helping to improve safety in the field [23].

E-learning and especially the implementation of m-learning, is being facilitated by the IoE across the educational establishment, giving more accessibility to students. The benefits include more feedback and monitoring of the progress of the learners [24].

## 4. Internet of Nano-Things (IoNT)

*4.1. Core Ideas of IoNT*

The concept of IoE is being extended to its fullest by the implementation of the IoNT. This is being achieved by incorporating nano-sensors in diverse objects using nano-networks. A model of this concept as a medical application is shown in Figure 6: this provides access to data from in situ places previously inaccessible to sense from or by the use of certain instruments that were impossible to use due to their former bulky sensor size. This will enable new medical and environmental data to be collected, potentially leading to the refinement of existing knowledge, new discoveries and better medical diagnostics [25]. The technology is described by Akyıldız and Jornet [26], using graphene-based nano-antennas operating at Terahertz frequencies. They also discuss the problems of extreme attenuation operating at these frequencies and networking at this nano-level [25]. Each functional task, such as actuation or sensing, in an IoNT is performed by a "nano-machine"—whose dimensions may range from 1 to 100 nm [22].

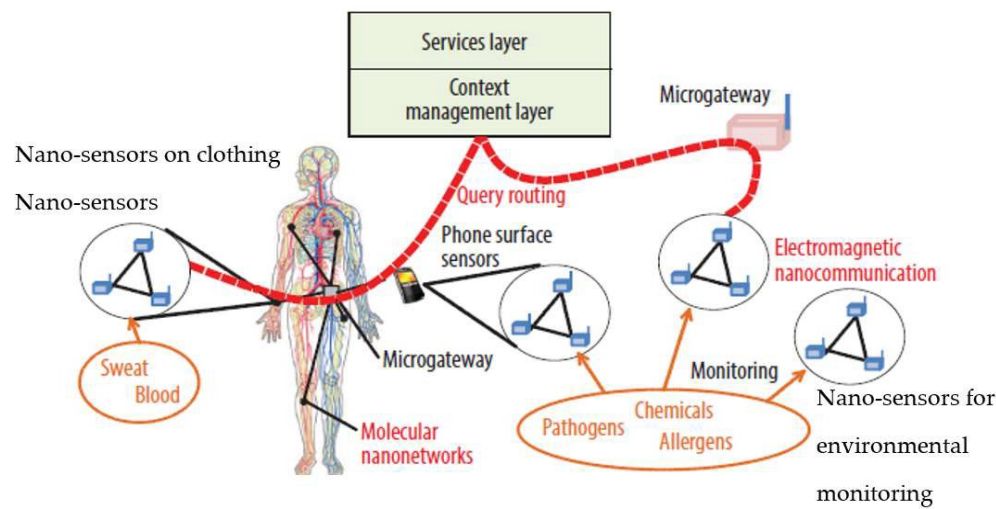

**Figure 6.** The Internet of Nano-Things. Adapted from [25].

Thus, the Internet of Things will not only be deployed in the world that can be seen, but at scales that are invisible to the naked human eye. This will be by the use of IoNT and IoBNT. Their use will not only be medical at the cellular level but industrial, for example in filtration work such as water purification or for dialysis. The overcoming of a major obstruction of IoBNT will follow from the seamless merger of IoNT with existing health-based IoT systems as well as networks [27]. The application of IoBNT, being stemmed from synthetic biology as well as the utilization of nanotechnology tools to enable the engineering of biological embedded computing devices [28], will reduce the risk of undesired effects on health and/or the environment.

*4.2. IoNT Future Trends*

IoNT devices, being in their infancy, are currently relying on the established protocols of the Internet. These will need to be adapted for the particular requirements facing IoNT devices, such as the challenges of the communication and power requirements of such small devices. These will obviously need to be solved but are considered to be within relatively easy reach of software and hardware developers. Implementing transceivers to demonstrate the practicality of IoNT has been shown to be theoretically possible, with research on graphene radios using Terahertz frequencies. Thus, invasive monitoring in situ can be implemented using implantable biosensors. They may also be used to monitor the environment, such as watercourses. Graphene-based transceivers have been shown to operate at one terabit per second due to the high bandwidth, but the ancillary electronic components to make the nano-transceiver a reality are still being researched [29].

The interfacing of IoNT with existing micro-devices is important for it to ever become all-pervasive—further study should be focused on this task, especially in the industrial, biomedical and industrial arenas. Major challenges need to be addressed in the fields of electromagnetic channel modelling at this biological cellular scale and the necessary supporting networking protocols[26].

**5. The Future Internet**

Based on the Gartner Hype Cycle of 2014 [30], Forbes [31] reported in August 2014 that IoT had overtaken Big Data as a topic of discussion, with over 45,000 references in the media in 2014, compared with only 15,000 in 2013. The Gartner Hype Cycle shows the lifetime of a particular technology from inception to maturity to decline; this is particularly helpful in mid-term business planning.

However, Gartner has retired Big Data in their 2015 report [32] since Big Data became truly prevalent and pervasive across many other hype cycles, such that it is no longer considered an emerging technology. As Figure 7 shows, according to their 2017 report [33], IoT still remains at its peak of hype, although it has experienced a shift in categorization from "five to ten years" to "two to five years" to reach maturation.

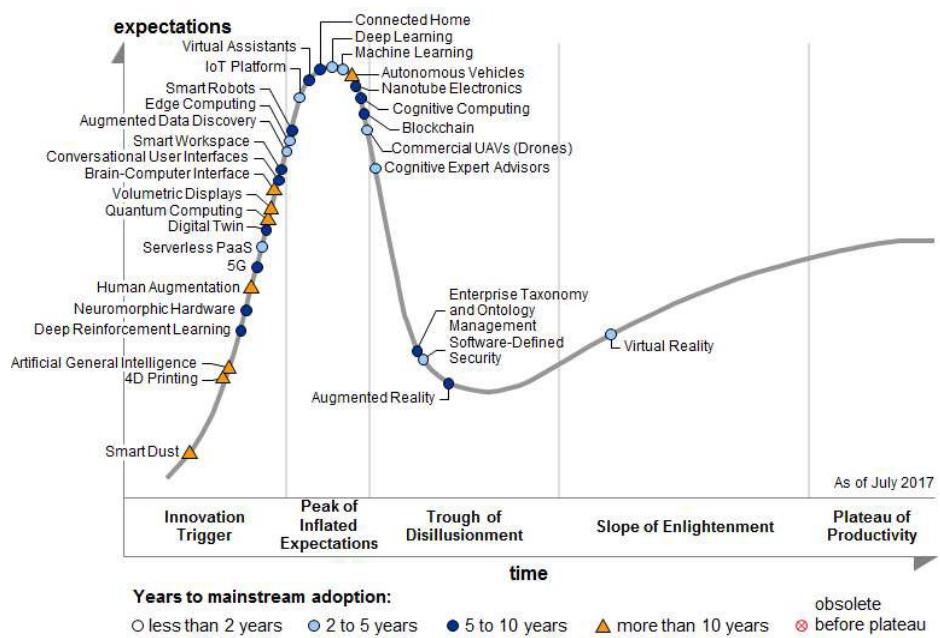

**Figure 7.** Gartner Hype Cycle, July 2017 [33].

Much research is being conducted in the field of IoT in the three domains of user experience, engineering and design [4]. The emphasis is particularly on the end user and accessibility. This is especially pertinent as 50–200 billion artefacts are likely to be internetworked to the Internet by 2020 [4].

To help achieve a more user-friendly interface, user-centered tools such as Microsoft's Gadgeteer may be employed [4]. This tool provides rapid prototyping of connected devices [34]. Theories from cognitive psychology [35] have also been utilized to design adaptive IoT systems. This technique relies on using the "FRIEND::Process" tool for human task organization and for both bottom-up and top-down organizations [4].

Simpler embedded devices will form a significant part of the future IoT. Many difficult financial, technical and social issues remain to be addressed [34], but the reality is that the IoT does now exist and uses standardized international networking protocols [36] with IPv6 forming its core foundational routing protocol [37].

For the objects that compose the IoT to acquire "ambient intelligence" they must comprehend the end user as completely as possible. This may be achieved by: observing, monitoring and recording the human users' body movements, gestures, location, context and environment. This will be likely to lead to high levels of user support requirements that were unknown previously in computing history [37]. The understanding of neuroscience, psychology and human behavior will thus play an increasingly critical role in achieving ambient device intelligence. The devices must use Artificial Intelligence (AI) to understand how humans process information and interact appropriately within the right social context and multi-user scenarios [35]. In fact, Ferati et al. [38] have demonstrated the feasibility of conducting software requirement analysis using IoT, especially for people with special needs: such people are likely to receive early and uncontroversial benefits from the technology and hence are a priority for development work.

The UK Open University offers users a course on IoT with programming and real-world sensing applications [39]. This is a first step in addressing the shortage of IoT engineers and programmers, especially as consumers become producers [39]. Educators will need to address many issues, not only the technical but also ethical and privacy issues. The Open University course was listed in the 2012 New Media Consortium (NMC) Horizon Report [40]: the report also predicted IoT adoption around 2016–2017. Hochschule Aalen of Germany [41] is now offering a dedicated full-time Bachelor of Engineering Degree on the Internet of Things, created with the aim of disseminating the broad technical knowledge of deployment of IoT sensors and their associated electronic hardware and software (programming for servers and big data). This practice-oriented 7-semester degree curriculum not only covers various technical aspects from electronics and computer science but also incorporates modules in business including, for example, IoT business model development. The degree focuses on user-centered design and development and the projects are interdisciplinary, enabling the real-world application of the acquired knowledge.

Research continues with the European SENSEI project, concentrating on the future underlying architecture of the IoT and its services [36]. For IoT to be a practical pervasive reality, it must be able to coexist and integrate fully with the Cloud. This means using the current Internet Multimedia Subsystem (IMS) platform to integrate both technologies [42].

Due to the successful deployment of various novel, innovative and useful applications based on IoT/IoE, the usage of computing devices and the Internet by people from different cultures, socio-economic backgrounds, nations, religions and geographical diversity is increasing at a near-exponential rate. As a result of these phenomena, universal usability or Ubiquitous/Pervasive Computing [43,44], Usability [45] and User Interface Design [46] have become very active topics and Cross-Cultural Usability [47] and Plasticity of user interface design [48] are important emerging areas of work. Exploring and analyzing the Cross-Cultural Usability and Information System (IS) issues [49–52], focusing on Web and mobile interaction using IoT/IoE as well as adoption trends and Diffusion of Innovations [53–56], are priorities to be researched in depth. These are important trends among users in how the "IS" is being utilized. As has been rightly pointed out by Ben Shneiderman, contemporary Computing is all about what users can do rather than what computers can do [43,57]. Thus, for the future, the success of the IoT/IoE must consider the impact of cross-cultural usability by intensive research in this direction.

The three major recent trends shaping the transformation of automation technology are: Tactile Internet, Cyber-Physical Systems (CPS) and IoT [58]. The latter two rely extensively on mobile Internet connectivity (i.e., telecommunication networks) for their typical operations, due to using solely wireless Internet-based communication. Thus, they were not highly adopted in industrial automation in the past since they could not be a means of efficient, reliable and deterministic communication for automation-specific requirements. However, improvements in reliability, battery power, energy harvesting and minimization of power demand mean that automation technology is now utilizing more and more IoT devices.

5G [59] mobile technology is designed to truly implement a heterogeneous network, which is just what IoT optimally requires. This is intended to cover both wired and wireless communication, both terrestrial and non-terrestrial in nature, including the use of IoT devices. The same stringent specification standards need to be adopted for IoT devices.

The use of fog computing [60] (also known as edge computing) along with cloud computing will greatly facilitate the use of IoT devices. Security can be implemented with the use of concepts taken from the decentralized blockchain [61–64] concept used in the Bitcoin cryptocurrency network. Two specific cryptocurrencies for IoT devices have already been designed and are being deployed, known as IOTA [65] and EOT [14].

Power sources pose a major problem with IoT devices, hence the need for energy harvesting. Novel solutions are beginning to emerge such as wireless powered communication networks (WPCNs), the energy required being obtained from a hybrid access point (HAP): this is termed a hybrid because both energy and information are exchanged.

## 6. Challenges and Impediments to IoT

As with any new technology, there is usually some inertia in the pace of its uptake. Currently the largest three impediments are due to technological factors and not human resistance, these being: standardization of protocols, global implementation of IPv6 and power needed to supply the sensors. The following is a list of challenges and impediments that IoT is currently facing or will face in the near future:

### 6.1. Deployment of IPv6

In February 2011 [66] the supply of IPv4 addresses held by the Internet Assigned Numbers Authority (IANA) was exhausted. The ushering in of IPv6 (Internet Protocol version 6) was critical to cover this IP address shortage, as billions of sensors will each require a unique IP address. The deployment of IPv6 will further make network management less complex, with its enhanced security features and network auto-configuration capabilities. However, the deployment of IPv6 has its own challenges, the following are the major probable ones:

- In its infancy, intruders, man-in-the-middle attacks or any general attacker may demonstrate a greater level of knowledge and expertise in IPv6 compared to the IT professionals, including the network administrators of any organizations. During the nascent period of deployment, it may initially be very strenuous to manage and discern unauthorized or even unidentified IPv6 assets within the legacy operational IPv4 networks.
- Operating both the protocols simultaneously during the transition period may also add to the overall complexity and cost in terms of time, human resources and monetary value.
- A prolonged period for IPv6 to mature, especially in terms of implementing it in security protocols and devices, poses additional risks.
- An increasing myriad of IPv6 tunnels along with the existing IPv4 ones, may add extra layers of complexity to the existing defense mechanisms.
- Another major challenge will be finding an optimized approach of dealing with the existing legacy systems, assets and devices.

To address these overall challenges, along with plans for a phased development, programs of education and training for IT staff to widen their knowledge and expand their expertise in IPv6 need to be seriously considered.

*6.2. Sensor Energy*

Due to the extremely adaptive nature of IoT-enabled devices, with their consequent wide-ranging and dynamic energy requirements, any IoT infrastructure must be energy-consumption aware to ensure its longevity of operation: this also affects its economic viability. It is estimated that just the US data centers will be consuming around 73 TW (terawatts) by 2020 [67]. This is enough to power two cities the size of New York.

Supplying reliable power to the sensors for a prolonged period is key to IoT being deployed successfully [5]. This is especially of major concern where these sensors are employed in remote and distant locations such as under the ground, in the sea, outer space or on other planets. The energy demand must be minimized and the supply must be harvested from the environment. Since it is not feasible to change the batteries for billions of these devices. Several technologies are being pursued to achieve this, including solar cells, thermal generators (using the Seebeck effect [68]), rectification of radio signals and exploitation of the energy in vibrations and other peripheral movements. One technology to facilitate this is the adoption of the concept of Narrow Band Internet of Things (NB-IoT) to implement a Low Power Wide Area Network (LPWAN). The use of Bluetooth Low Energy (BLE) transceivers has also helped in the deployment of IoT devices.

*6.3. Standardization*

Foremost in addressing the latest requirements for the pervasive implementation of IoT in terms of meeting stringent privacy and security requirements and at the same time adopting an elastic network architecture [69], is the work of the IEEE standardization organization—especially in regard to adoption of IPv6 packet routing through increasingly heterogeneous networks [5].

Because of heterogeneity in networks as well as devices, interoperability is a fundamental need for the functioning of the Internet. This is more obvious for the IoT ecosystem since billions of devices are connected to the Internet as well as to each other. Each IoT device should converse in the same language (protocol) so that other devices can understand and thus standardization is paramount. The so called "walled gardens", providing a closed platform or ecosystem, which only allows communications limited among the devices belonging to the same vendor, restricts the advantages of having Internet access. Because IoE comprises multi-vendor devices, IoT systems need to go through intense and rigorous interoperability and compatibility tests before they are formally launched. This, however, does add to extra complexity and cost.

While standardization is still in progress, constraints such as cost, time to market and technological limitations faced by IoT device manufacturers also adversely result in poor interoperability, lack of conformation to standards and often a compromised design. Technical and technological constraints include having to deal with limited memory and lower processing power as well as power consumptions having to be satisfied by limited non-renewable power supplies. Moreover, industrialists are under the compulsion to minimize the unit cost and thus the overall product costs to maximize profits. It has become a norm to use cost-benefit analysis, which may indicate it to be economically attractive to trade-off interoperability and performance (with their additional costs) against short-term goals of producing cheap IoT devices (potentially non-secure), even sometimes leading to proprietary products. However, approaches should be implemented to compel manufacturers to actively consider international policies such as adopting interoperability and conformance to international standards that yield long-term benefits, including product life-cycle gains.

*6.4. Architectural Limitations*

The adoption and viability of IoT clearly puts many technological demands on the current Internet infrastructure. Many of these limitations have been clearly exposed when IoT devices were practically implemented over the current Internet infrastructure, such as: weak security, interoperability problems, data provenance and excessive human interaction. These problems still require to be addressed for the rollout of 5G and for the deployment of the Future Internet (FI).

Because IoT devices are composed of so many different technologies, when networked, they form inherently quite complex structures. Thus, IoT network failure may require more time for fault diagnosis and restoration of service. This also means having maintenance personnel with multiple networking and protocol skills. This naturally entails a more expensive workforce to both hire and retain.

The architecture of an IoT system may be broadly classified into four layers, as shown in Figure 8, below.

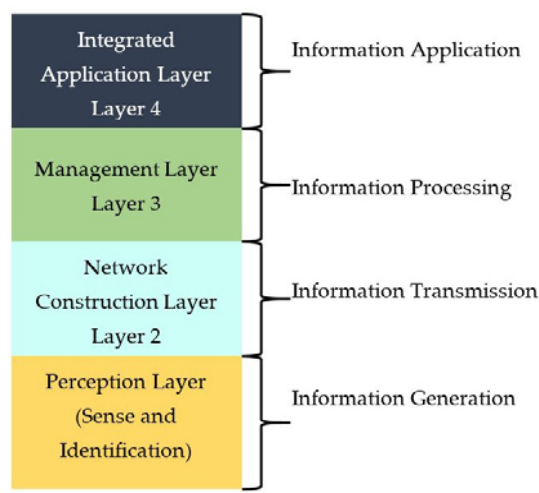

**Figure 8.** Four-layered architecture for IoT.

Constructing a global unified IoT ecosystem that is communicating transparently is still not possible yet, at the time of writing of this paper. This is due to no universal protocol currently being in place that can work across heterogeneous networks. The IoT at the moment is just a collection of Intranets of Things. For IoT to be a seamless ecosystem, a standardized internationally agreed application layer protocol needs to be created. This protocol would also take into account communication across the various physical interfaces of the IoT devices. Instead of inventing a new protocol, it would be far easier to reuse the technologies from the Web itself. The "Web of Things" [70] has precisely these goals needed to make the IoT ecosystem a reality. These are to: "reuse and leverage readily available and widely popular Web protocols, standards and blueprints to make data and services offered by objects more accessible to a larger pool of (Web) developers." [70] The Web of Things does not actually stipulate the physical layer connections between devices, thus the Web of Things will function whether connected to a company intranet, domestic network or any type of LAN. To clarify, the Web of Things (WoT) encompasses every aspect of the software approach to make WoT be fully integrated into the World Wide Web (www). Analogous to the mapping of the Web (Application Layer) to the Internet (Network Layer), the WoT also has an Application Layer that aids in the authoring of IoT applications.

*6.5. Pervasiveness*

The total number of connected IoT objects is projected to increase from 21 billion (2018) to over 50 billion by 2022 [71]. This clearly illustrates the spread of truly pervasive computing devices and the challenges they will have to face in their overall management. Thus, IoT devices will need to be autonomous for their successful deployment, with little or no human intervention at all. This prerequisite, along with the ubiquitous nature of IoT, raises trust, security and reliability concerns, especially if utilized in the healthcare sector. An example would be the use of IoT in critical life support systems: such concepts also raise several very significant ethical concerns.

Furthermore, to process the vast amount of data being collected by IoT ubiquitous sensors and the specific needs for this to be processed, big data analytic techniques need to be considerably enhanced.

The pervasiveness of IoT devices is driven by their exponential adoption rate and this shares similar concerns with the field of pervasive computing. Some might argue that IoT is more concerned with the realm of device connectivity, whereas pervasive computing deals with human-computer interaction (HCI) matters. However, they both share common technological issues, such as ensuring security, privacy, ethical behavior and common applications. Thus, it would make sense if both communities worked together as proposed by Eblings [72].

*6.6. Retrofitting IoT Devices*

Retrofitting of IoT devices with additional sensors is not easy once they have been deployed, particularly if they are inaccessible in a hostile environment. Thus, multi-sensing sensors should be utilized to overcome this logistical problem. One solution is to use backscatter-enabled passive sensor tags that add new sensing capabilities to IoT devices in their near neighborhood [73], as shown in Figure 9.

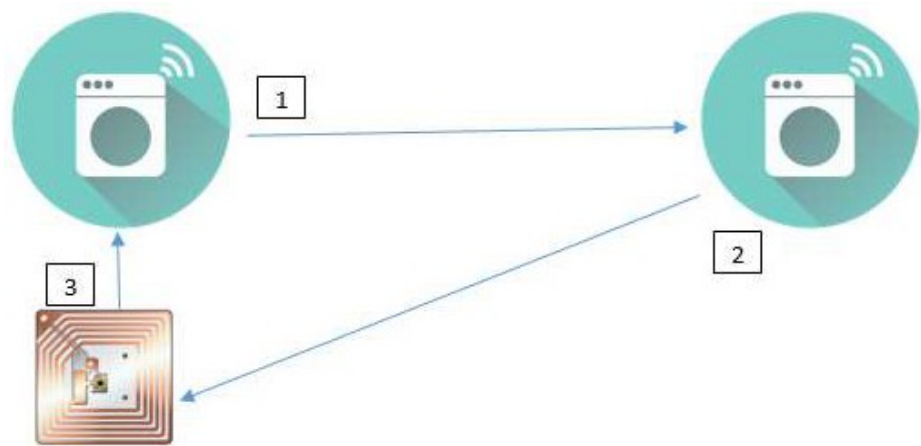

**Figure 9.** Passive sensor tag placement near other IoT devices to add extra sensing capabilities.

Referring to Figure 9, Device (1) does not have the requisite in-built sensor. However, the required sensor (3) is located nearby. To query the passive sensor tag (3), device (1) transmits an unmodulated carrier to its neighboring IoT device (2). The passive sensor tag, device (3), upon receiving this request from device (2) then modulates the carrier with a valid 802.15.4 packet. This is then transmitted to the requesting device (1). In this scenario, it is assumed that device (1) cannot send a signal directly to device (3) to generate the valid 802.15.4 packet.

*6.7. Multifaceted Exponential Growths*

The Internet is adapting to new services with the use of different protocols specifically to support IoT devices, such as the use of IPv6 over low power wireless personal area network (6LowPAN) as an example. Radical and revolutionary approaches will be necessary, such as redesigning segments of the Internet infrastructure dealing with IoT devices. It should be noted that these segments may consist of millions or even billions of devices. The 5G architecture and FI are being designed to both cater for and address this.

The key to supporting IoT devices is the ability to handle disparate requirements such as some devices requiring no security while others need to be highly secure; devices with low data rates versus those producing very high data rates; critical (e.g., medical) data demanding high priority for timely transmission, etc. Thus, a carefully designed resource allocation strategy will remain as one of the core concerns. The concepts from the Cloud infrastructure and services can be taken into consideration in meeting these challenges.

*6.8. Software Defined Networks (SDN)*

SDN offers the flexibility and adaptability needed for the successful universal deployment of IoT devices. The integration of SDN with IoT is thus considered to be another major and urgent challenge. Cost constraints and the time to market will also influence the operability and design of IoT devices. The challenges presented demand the adoption of universally agreed upon standards for the IoT devices to operate successfully in the global market.

The fluid malleability of Software Defined Networking (SDN) is seen as being a positive disruptive force in computer networking. Benefits include programming network switching elements (forwarders) to program packet routing to any port based on any specified packet parameter. This is of special concern that will benefit IoT devices to communicate with each other over a heterogeneous network. Thus, amalgamating SDN with IoT is one strong way forward. However, this poses other challenges that must be circumvented, especially in operational validation of the hybrid-combined network against the current disparate networking solutions.

The adoption of SDN is accelerating because of its huge potential in hardware cost savings. SDN through its very nature of extreme software configurability allows generic network components to take on any function, such as a switch or router. SDN can be seen as the stem cell programming of computer networking. The SDN model also differentiates and isolates the control and signaling plane from the data planes. The intermediate networking elements of SDN have also been simplified to be forwarders of packets. A general control protocol is used with forwarding rules to achieve this. A central processing unit or the brain of the SDN also acquires the stable topology of the SDN network. This then allows optimized routing decisions to be made along with optimized and parsed forwarding rules. Adopting the characteristics of the SDN can be utilized to deal with IoT operating in a heterogeneous network. Thus, the IoT network can be scaled and a new high-level control solution can be created that interfaces seamlessly with the SDN controllers, as shown in Figure 10 [74].

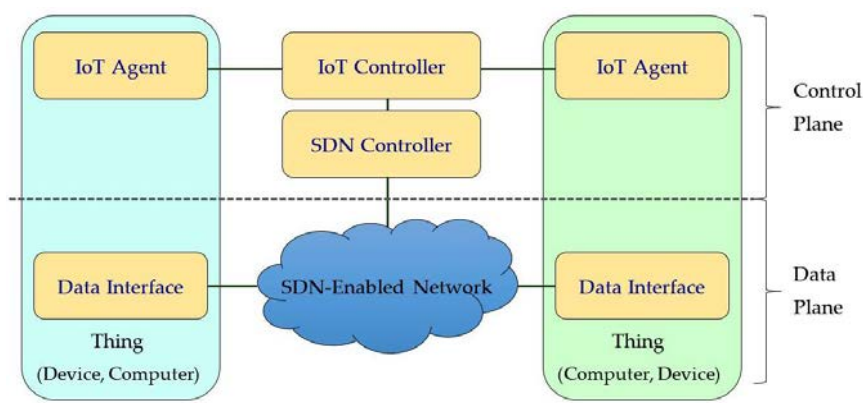

**Figure 10.** IoT and SDN Conceptual Integration Overview. Adapted from [74].

As shown in Figure 10, there exist two distinct planes, the control plane and the data plane. Two IoT objects can interact with each other through the SDN-enabled network using their respective internal IoT agents. Contextual information is conveyed to the IoT Controller, which then passes that to the underlying SDN Controller. The IoT Controller, though shown as a monolithic block, may in fact be composed of several internal modular blocks. This flexibility enables new functionalities to be added to the IoT object without affecting its final relation with the SDN Controller.

The major current impediment for a universal model as shown in Figure 10 is the lack of a stable IoT architecture. The other major factors delaying the creation of the universal model is that there is no standardization for IoT content awareness provisioning and Quality of Support (QoS).

*6.9. Fog Computing (Edge Computing)*

Fog computing, as coined by Cisco, is a particular form of cloud computing, primarily differing in terms of the location of operation. Fog computing, also commonly known as Edge Computing, being operational at the edge of the enterprise networks, extends the cloud computing system. Due to its location being close to the edge of the enterprise network, fog computing provides comparatively better performance in terms of reduced delay, lower latency and jitter—which in return result in improved processing time and near real-time responsiveness of networked user applications. Furthermore, it provides other location-based advantages such as better customization options as well as mobility support. Because the fog computing approach requires the decentralization of a major share of the complete data processing components (such as applications and services, computing and processing power, data analysis and decision making) as well as the data itself, it lessens a substantial volume of network traffic flow, especially in data transmission between the Cloud and IoT devices.

Cloud computing is not redundant, however, as batch processing jobs are still heavily generated by the scientific and business world. These types of computing tasks are best processed in the Cloud. Fog computing is ideal where data processing needs to be executed at the point of its generation. Thus, data analytics and knowledge processing and generation can occur at or near the data source. Fog computing, due to its localized nature, also allows for better locally optimized applications. To meet future performance requirements of integrated IoT and Fog Computing, the architecture of Fog Computing needs to meet the strict tolerances and requirements of energy savings, data throughput and latency constraints at both the node and system level. Hence, Fog computing still needs to evolve further to meet performance requirements of IoT over the use of contemporary Fog nodes.

*6.10. Limitations of Current Wireless Sensor Networks (WSNs)*

To be able to securely manage and control embedded IoT devices is a key functional requirement for their successful operation. An efficient and optimal architecture for secure software update/management thus needs to be designed. Furthermore, the current approaches related to WSNs need to be specifically tailored for IoT devices. This is so because they are not well adapted for the operating requirements of IoT devices, such as in power requirements or energy-aware routing needs. A system level view is shown in Figure 11 of a WSN [75].

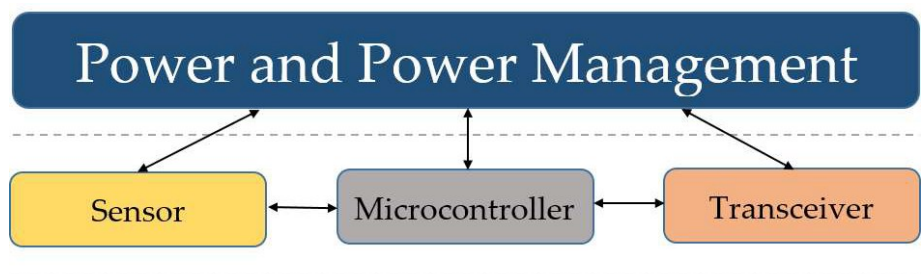

**Figure 11.** WSN Sensor Node Hardware Structure. Adapted from [75].

A WSN is typically composed of a cooperatively sensing network of nodes. These nodes may both monitor and alter the environment and interact with humans. This is best served by a cross-layer design approach requiring distributed processing, communication protocols and MAC (media access control) querying. IoT will need to be able to coexist with many different wireless and wired technologies, including integration with WSNs. Thus, for IoT to be truly pervasive, WSN will form a critical component of IoTs, featuring low power requirements, ruggedized design and low price. WSNs also need to be massively scalable, including the associated requirement to fit into a system that can handle intelligent control, massive heterogeneity, dynamic service changes, concurrency, real-time operation, enhanced security and multiple access techniques.

*6.11. Ethical Issues*

IoT devices are expected to permeate the whole fabric of our socio-technical ecosystem, including not just the implementation of Smart Cities, but also invasively within our bodies for total healthcare. This naturally raises many ethical issues that need to be resolved to allay the concerns of the public. Protection from eavesdropping of medically sensitive data is paramount to protect patient confidentiality. Also, protection from hacking, which would have particularly disastrous consequences for medical monitoring and equipment maintenance, must be guaranteed to safeguard life.

Any automation of manual labor risks employee redundancy and hence potentially massive job losses. This is particularly so for unskilled and low-qualified blue-collar workers. Examples of job losses are in the sectors of inventory and stock control, check-out machines in stores and ATMs in banks that now do more than just dispense cash. The use of IoT may consequently lead to the widening of the gap between the rich and the poor. Those particularly disadvantaged are the impoverished and those lacking or with no access to the Internet at all. This is specifically acute for the less industrialized nations and similar regions within an otherwise industrialized country.

The erosion of privacy may be seen as inevitable with the adoption of the IoT; however, the professional institution ACM (Association for Computing Machinery) states that it will respect privacy and honor confidentiality. Already the habits of consumers are being collected in minute detail. The granularity of the data collected is expected to get even finer and more intimate with the adoption of IoT and smart devices in the home. Data mining of consumer behavior from diverse and disparate sources allows advertisers and marketing agencies to build up a complex and very rich profile of the consumer. This makes targeted advertising even more relevant to the consumer, an example being that algorithms can even predict whether a female shopper is pregnant before her own awareness. Data mining with data from more and more IoT devices, if uncontrolled, can only help to increase the accuracy of prediction of the short and long-term behavioral patterns of the consumer: as is now well-known, this can extend beyond the commercial domain and into politics as well.

Data provenance, including the identity of the "creator" of the data and the rights of this "creator" need to be clearly identified from the beginning. The creator could be IoT devices themselves, in which case the legal entity constituting the owner of these devices needs to be clearly established beforehand. This is particularly imperative when dealing with financial transactions. Here, another revolutionary technology, the blockchain [62], may help immensely. Thus, the use of data provenance and the blockchain may be utilized to clearly delineate the private domain, public domain and the personal domain of IoT environments [76]. Personal information, as collected especially by IoNT, BIoNT, MIoT, CIoT and HIoT, covering medical information in particular, needs to be strongly protected from malicious use and hacking.

It is important to ensure protected accessibility to information to safeguard it from virus attacks, hackers and consequent information loss and spoofing as all such attacks may adversely affect the lives of people. A car connected to the Internet may need to be strictly protected from malicious access, as it could be used to cause accidents and kill its occupants or others. Examples of industrial espionage and sabotage include the computer worm "Stuxnet", which was used to carry out a cyber-attack on Iranian power stations, as reported by Sky News [77]. However, the grandiose claims by Sky News that Stuxnet could be used to attack any system connected to the Internet, such as utility companies, hospitals, distribution networks, traffic systems and heating/cooling systems were later debunked by Sophos [78] and Eset [79] as "over the top reporting". Nevertheless, such an example of a primitive cyber-attack for malicious purposes may be seen as just the beginning of more advanced future destructive warfare upon a nation, to parallel conventional warfare.

The digital divide is likely to grow with the Internet of Things, as it will only be able to be fully exploited, deployed and utilized by countries having a substantial technically competent skilled workforce and management. Those nations with the resources to train and educate new security experts will thus have an immediate advantage over the unprepared.

How communication between the various IoT devices will impact human lives is of particular concern, covering not only the psychological factors but the legal issues of privacy and human rights. The massive interconnectedness and pathways of communications between IoT devices raise particular legal and ethical questions covering the:

- Privacy of Information
- Data Ownership
- Ethical and Legal Usage of Sharing of Data
- Security of Information Flow and Storage
- Transparency of Data and Data Provenance
- Data Collection Rights and Protection from Nonfeasance, Malfeasance and Misfeasance
- "Digital Knowledge Divide" and Minimization Thereof.

One of the major drivers of "Industry 4.0$^t$ (cyber-physical systems) is automation of the complete production ecosystem [80]. One school of thought holds that this is presently causing an increase in the unemployment rate, predicted to rise exponentially in the near future—when more industries adopt Industry 4.0. The debate regarding the ethics of giving human jobs to machines is centuries old and has now moved on to discussion of robots and IoT devices However, the concept of "Industry 5.0" is ready to emerge which aims to return human hands and minds back into the industrial framework. If adopted, Industry 5.0 will thus ameliorate the massive layoffs envisioned by the adoption of Industry 4.0. The five stages of the industrial revolution are shown in Figure 12, below.

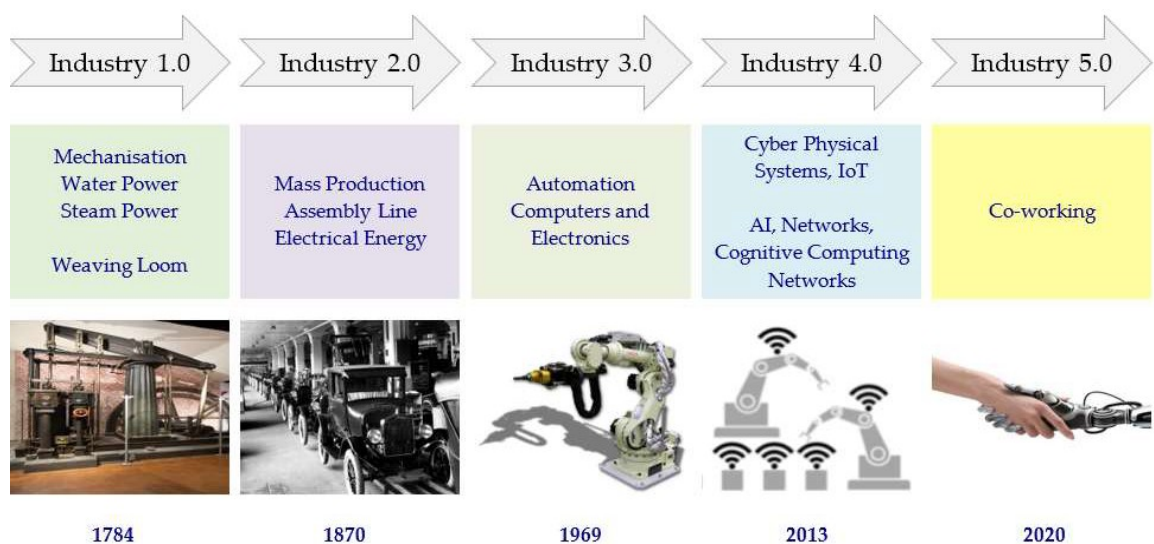

**Figure 12.** The Five Stages of the Industrial Revolution. Adapted from: [81].

*6.12. Vulnerabilities*

Scanning of the entire IPv4 address range back in 2012 found that 450 million devices were wide open and accessible [82]. This is clearly a major vulnerability issue that needs to be closed if IoT devices are to be made secure and illegal access to them denied. Recent (2017) demonstration of breaking of the Wi-Fi Standard WPA2 security protocol by key reinstallation attacks (KRACKs) forcing nonce reuse [83], proves that Wi-Fi-based IoT devices are at risk too. The ways to increase protection of Wi-Fi security are explained in [84]. Mobile Internet connectivity being more secure than Wi-Fi and with the deployment of 5G, the IoT industry may move away from Wi-Fi to 5G for better connectivity and security. The use of the Blockchain [76] also offers the potential to implement a more secure and private IoT ecosystem. Furthermore, the use of statistical fingerprinting may aid in developing a more secure IoT world—which is essential for the healthcare sectors [22].

The hacking of most IoT devices is currently relatively straightforward due to their complete lack of any security features. By utilizing literally a myriad of insecure devices, a DDOS (distributed denial of service) attack can be executed to bring down complete infrastructures. This can have disastrous consequences for human society. Another example of an attack is using just one weak IoT device to gain entry as a gateway deep into a network to possibly access sensitive, critical and valuable data. The largest DDOS attack in human history so far occurred in October 2016 using an IoT botnet to spread the Mirai malware, which caused large sections of the Internet to go down. The Internet service provider Dyn was targeted and major services including Netflix, Twitter, CNN, Reddit and The Guardian all experienced service disruptions. The Mirai malware searched for those devices whose usernames and passwords had not been changed from their default values to infect them. Digital cameras and Digital Video Recorders (DVRs) were not immune from this attack either. The lessons that may be learnt from this particular incident are these [85]:

- Do not use any devices that cannot have their usernames, passwords, drivers, software and firmware updated.
- Change the default login details immediately on acquisition for any Internet connected device.
- Each IoT device must be assigned a unique password.
- All IoT devices must execute the latest firmware, driver and software to protect against security vulnerabilities.

*6.13. Privacy Issues*

The IoT ecosystem can be considered as an abstraction of the real world, hence a form of virtual reality, representing detailed monitored real-world events in the digital realm. Consequently, legal and ethical issues, as well as concerns allied to data protection law and privacy, remain identically valid. These need to be equally considered in the same manner as security and technical issues. Protection and storage of the vast amount of data generated by IoT devices, e.g., using the Blockchain [62,76] and Big Data [20] concepts respectively, need to be meticulously engineered to ensure the utmost privacy.

Sensitive data collected by smart IoT devices need to be protected against privacy violations by careful management [86]: currently most devices do not offer any protection. Location information needs to be protected along with its associated metadata to prevent malicious access. This is especially important to stop widespread fear among the public, potentially preventing the wide-scale adoption of IoT. The fear of "Big Brother-like entities" is discussed in [87]. This major issue of IoT privacy must be resolved effectively by the IoT community to allay fears from the general populace. [88] also discusses the potential problems that may be created by self-aware IoT devices including in general the following IoT issues: "data integrity, authentication, heterogeneity tolerance, efficient encryption techniques, secure cloud computing, data ownership and governance, as well as policy implementation and management" [88]. [87] offers some solutions, such as: building the IoT device with designed inherent privacy; user defined data management and access rights. Data flow transparency is also suggested so that users of IoT devices know exactly who has their data [86]. Complete data management, taking into account policies and the use of enforcing instruments, is proposed by [89], which also discusses the need to adopt the attribute of typifying the data, its ownership details, the span of access, its viability and anonymity.

Users should also be offered the option to "opt-out" [90] if any sensor is deemed to be an untrustworthy node. This is also known as "right to silence of the chips" [91]. Proxies may also be used that act as a "privacy broker" [92]. The socio-economic and ethical aspects of the usage of IoTs need to be addressed, as pertaining to privacy and not just the technical solutions. This will require education of the users of IoT devices so that they know how their devices gather data and process it as well as updating the current privacy regulations. This is quite an onerous task, requiring determination of the distinction between IoT Personally Identifiable Information (PII) and regular information. The question of whether these regulations should be dictated by governmental agencies or the current self-regulatory agencies is still undecided. The scopes of these regulations also need to be established to cover their territorial jurisdiction and collaboration with civilian partners. The European Commission and the US Federal Communications Commission (FCC) have already started work on their recommendations [86].

*6.14. Automatic Discovery of Resources*

Due to the wide diversity of IoT objects and devices, a universal protocol for the automatic discovery of resources must be created. This IoT application-specific protocol cannot, however, be just a simple modification of SIP (session initiation protocol). In traditional computing the end users are more aware of the software and resources of their computing devices. This is not the case with IoT devices, which, being autonomous, must carry out automatic discovery of available resources, without human intervention. This is particularly acute for the millions of IoT devices likely to be deployed in Smart Cities. This approach is in direct contrast to mobile platform applications usually under the full control of the user.

To aid the automatic discovery of resources and to help the IoT end user, different value-added services can be created. These services can extend beyond the normal IoT daily functions to include: semantics and data identification services, automatic configuration management, device registration/deregistration, service advertising and device semantic integration. Thus, for this process to be a complete success, IoT objects need to be discoverable and allow themselves to be discovered within the IoT ecosystem, with full mutual exchange of device capability information. Device prioritization also needs to occur, hence implying the ranking of IoT objects along the IoT network chain. This entails how far and wide an IoT needs to be detectable to make its presence known to the network initially. The discovery process can be event-based, one-time only, on a publisher-subscriber basis and indicating whether it is a home device or an IIoT device.

Resource discovery, though essential for an IoT device, needs to take into account the power consumption from the limited IoT power supply. Thus, IoT devices, through their power conserving nature are mostly in a dormant state and only "wake-up" when required, e.g., an IoT fire sensor only alerting the user via the Internet upon actual smoke detection: in the dormant state, this IoT fire sensor will be asleep and undiscoverable via the web. CIoT devices in the home are often behind a fire-walled gateway, thus they are not discoverable by web crawlers.

IoT devices may also be connected by low power radio links, often using shared electromagnetic spectrum suffering high levels of attenuation, interference, multi-path effects and distortion. This often means loss of connection and hence multiple attempts to reconnect during a session.

*6.15. Identity Management of Connected Devices*

With billions of devices expected to be in operation in the immediate near future, both security and device identity management are critical. A universal identity management scheme is suggested to resolve the issues of global interoperability, security and deployment [93]. To keep track of all these devices in this "identity ecosystem", computer scientists have begun to refer to this ecosystem as the "Identity of Things" (IDoT). The IDoT describes the realm of complex and cross inter-relationships between devices and with, humans, applications and servers.

*6.16. Evolution of Communication from H2H to M2M*

Communication has evolved from human-to-human (H2H) interaction to that between machine-to-machine (M2M), especially in the domain of the IoT universe. In fact, H2H, particularly voice communication, was the focus of early communication technologies. Thus, the existing network infrastructure, architecture and protocols are optimized mainly for human-generated data/traffic. This poses a challenge to the successful implementation of IoT, which can be overcome by the adoption of a new set of protocols to specifically support M2M communication as an alternative to those used just for digitally conveying the human voice.

Effective global H2H communication can coexist with M2M communication throughout the intermediary link. With IoT devices now outnumbering human operated devices, M2M communications naturally is the dominant and fastest growing technology because it also enables human contact as the ultimate end users. Because of the requirements of network resilience, fault tolerance and redundancy, an even greater number of M2M devices are being deployed, leading to its exponential growth. Device-to-device (D2D) forms the backbone of M2M communication [94]: this uses all available means of the global communications network to carry out its tasks.

*6.17. Need for Secure Data Management and Processing Solutions*

IoT systems are highly diverse in terms of functionality and applications and are also heterogeneous in nature. IoT-based applications are thus almost limitless in nature, ranging from wearable devices to distributed sensor networks. With all the wide-ranging types of data being generated, a unified, efficient and secured data management and processing strategy must be adopted for IoT to be operationally successful.

It is predicted that by 2020, 44 ZB [95] of data will be generated: 1 ZB is $10^{21}$ bytes. This vast amount of data that is being generated by both humans and devices needs to be processed and stored efficiently. This will require new high-density storage devices and, potentially, quantum computing.

The continuous generation of data, by interlinked IoT smart devices, needs to be controlled. The solution to this surfeit of data is yet to mature. Traditional big data, relational database management technologies and NoSQL (originally referring to "non SQL" or "non-relational") need to be scaled somehow as they are not adequate, if planetary IoT deployment is to be completely successful. Consequently, the present emphasis on just sensor IoT networks, need to be expanded.

The nature of IoT data and its frequency of generation need to be taken into account. This will require new mathematical studies and models to be built. The type of IoT data is likely to be intermittent, massive, geographically dispersed and often streamed in real-time [96]. This will require a complete overhaul of the present network components that make up the Internet to keep latency and jitter to a minimum. Not only the actual data need to be stored, but also its associated metadata. The metadata typically would consist of object identifiers, time and location of the data, services rendered, and processes occurred.

The nature of the IoT data will also vary as it traverses the network. It will travel through various types of flexible schema databases, fixed and mobile networks, concentration storage points etc. before reaching its destination via centralized data stores. This again highlights the criticality of successful management and processing of IoT data.

The newer models of databases will have to be adapted for IoT data to process: the remote storage of data at the "Things Layer"; the structure-less data; its non-atomicity; its less rigid consistency; its lower isolation and its lesser durability. This will be necessary, particularly in the drive for data availability and energy efficiency [97]. Also, data management must take on a dual role of offline storage and online-offline operations due to the dynamic nature of IoT-generated data. IoT data that is being generated needs to be summarized online along with the metadata attached. Further, the power requirement for each stage of the data generation and processing cycle needs to be studied more closely to optimize IoT power usage and extend the device longevity.

*6.18. Need for Big Data*

Since the source of data in IoT has evolved from human-to-machines to intra-IoT devices, the volume of such data is growing at a faster rate than the number of connected devices. This above-exponential growth in the volume of data now requires the use of Big Data architecture and data handling techniques.

The solutions offered for IoT data management are diverse and have not yet matured. Even the three concepts of Big Data of volume, variety and velocity [98], need to be tailored to deal with IoT data [20]. The nature of IoT data is such that: the volume of data will span from a few bytes to Gigabytes; the data will be very diverse, and the period may range from milliseconds to months.

Processing of such data allied with the use of artificial intelligence may offer better customer habit analysis with the aim of offering better services and experiences. In the scenario of smart cities, analysis of the data will enable a more efficient future city to be administered [19]. This can cover better traffic management [19], pollution control, utility services and habitation planning.

*6.19. Database Requirement*

Due to the diversified application of IoT in widely ranging domains, IoT data has the characteristics of being large in volume and integrally multidimensional, thus requiring frequent updates/writes. Although traditional Database Management Systems (DBMS) offer rich functionalities with multi-attribute access efficiency, they fail to scale-up to meet the increasing demands of high insert throughput and the sheer volume of IoT-generated data. Although Cloud-based solutions have good scalability, they suffer from not having native support for multidimensional data structure access. An example of a modification of Big Data analytics to support IoT devices is in the use of the Apache Hadoop database (HBase): which is based on an update and query efficient index framework (UQE-Index) [99].

*6.20. Modelling of Services*

Modelling of services and their interaction is key to the successful deployment of IoT devices, since, in contrast with the past, multifarious large industry software systems or applications are built from modelling services. Emerging technologies however, including IoT and CPS, pose more challenges since they need to be seamlessly integrated into already established existing models.

The backbone of IoT is based on the Internet infrastructure and the offer of real-world services. With advances in the Internet through such technologies as the SDN this also helps spread the IoT. The key to the success of the wide-scale adoption of IoT is the ability to provision real-world services. This will entail seamlessly communicating with heterogeneous objects. The sensed data from the physical world needs to be filtered and matched to precisely defined applications. Data fusion needs to be carried out from the collected disparate IoT sensors and the information presented in a meaningful manner. The decision making may be augmented by artificial and cognitive intelligence supporting autonomous reasoning. Further research needs to concentrate on the middleware to support all these new approaches and algorithms and seamless integration with the application layer.

The "Time to Live" (TTL) parameter from Internetworking is carried over to IoT data, combined with semantic modelling, annotation and metadata. Negotiations between the various IoT devices need to be carried out as fast as possible to reduce network delay. These device negotiations are necessary to discover what capability and services they can offer. Thus, IoT network discovery needs to reach a stable state as fast as possible despite the number of devices. Semantic modelling will need to take into account the unique constraints of IoT devices, namely their limited electrical power, memory and computational ability (noting, however, that these are expected to evolve in accordance with Moore's Law). The semantic modelling of IoT networks is carried out on powerful machines, such as the gateway nodes and in the middleware. Thus, the computational burden is shifted away from the IoT device, so that they can concentrate on gathering their sensed data. This concept of shifting the computational burden of those intensive tasks is similar to the approach adopted by the mobile communications networks. This also allows IoT devices to be queried more efficiently by software agents. Differing modes of processing and communication links will need to be utilized depending on where the data is in the network. Around the IoT device, low power and low bandwidth links will be required. Data gathered by data fusion can be processed in the middleware. Automation of all processes is essential, including making the IoT device as autonomous as possible. This is particularly important when it comes to the manual annotation of IoT devices used in semantic modelling: for a few devices in a controlled environment, manual annotation is possible, but as the number of devices increases this will clearly become impossible for a human operator.

*6.21. Notification Management*

Notification Management will need not only to monitor mobile communications from such services as the 5G mobile communications network but also from the IoT devices themselves. Hence, notifications will need to be prioritized and sent to the user in such a way as not to overburden the senses of the human user.

It must be stressed that the environment that the IoT devices operate in is very dynamic and hence the data collected and generated will also be dynamic in nature to reflect this. The dynamicity also extends to the software, drivers and firmware needed during the lifetime of the IoT device. Third-party applications need to fully take this into account and be able to meet this stringent need for IoT devices operating in a resource-constrained environment.

Service demands may also change for any IoT device. Their impact on the overall IoT network must be minimized. Thus, all the interfaces forming the IoT ecosystem need to be rigidly defined as internationally agreed and ratified standards.

**7. Concluding Discussions**

*7.1. Conclusions and Discussion of Future Trends*

As far back as 1984, the futurologist Ray Hammond, in his "The On-Line Handbook" [100], accurately foresaw that the linking of computers (i.e., the computer network and the Internet that we are using today) from all over the world would have far reaching effects, including: (1). The spread of knowledge; (2). The interchange of ideas and (3). The dissemination of information. Although he rightly further predicted that these were likely to bring a revolution in society, it is extremely difficult to precisely determine where the current developments in mobile applications, computer vision, consumer electronics, Artificial Intelligence and so on, mediated by the IoT, IoE and IoNT will lead us. However, Henry Jenkins [101] has offered an insightful explanation of the recent changes due to the digitization of media contents and their future impacts. We may expect to experience a period of transition for novel interactions, ubiquitous computing, mobile and ambient intelligent applications and the like, also mediated by Io(X)T, in the remainder of the 21st Century, paralleling that which was observed for personal computers and other similar devices, mediated by the Internet, during the latter part of the previous century. Although it cannot be guaranteed whether "Digital Immortality" as one outcome of the "Technological Singularity" can be achieved by the year 2045 or not, as forecast by futurist Ray Kurzweil in his famous book "The Singularity is Near: When Humans Transcend Biology" [102] or whether 'life' of "The World in 2030" will be "unrecognizable compared with life today" [103], but it is a truism that our life is increasingly becoming digitalized with the progressing inventions and adoptions of new technologies. Despite some negative aspects of this technological evolution, we can be optimistic about the coming computer revolution as technologies are becoming more affordable, convergent and novel in their solutions. Certainly, many tedious tasks can be taken over by linked inanimate objects and better availability of information must be a good thing. However, the dangers of criminal and other nefarious activity, plus those of hardware and software errors, pose major challenges.

Understanding and interpreting these trends is strongly dependent on insights in classifying different aspects, such that links between those that are similar are clearly identified but differences between those that merit differentiation are also identified. In this connection, the distinction between IoT, IoE and IoNT is seen to be a helpful differentiation that should aid insights in prediction of the near future.

The current prominence and future promises of the IoT, IoE and IoNT have been extensively reviewed and a summary survey report presented in this paper. The paper explains the fundamentals of IoT, IoE and IoNT and presents the recent research and advancements. The paper distinguishes between IoT and IoE which are wrongly considered to be the same by many. The discussion on IoNT presents limitations and challenges of IoNT by examining examples of previous research advances in the field: the concepts have been augmented by the inclusion of the IoBNT. The deliberation on "Future Internet", advocated by IoT, has been presented to reflect new research, associated challenges and future trends.

*7.2. Conclusions for Future Research Directions*

Upon examining the current advancement in the fields of IoT, IoE and IoNT, the paper identified and addressed the 21 most significant current and future challenges as well as scenarios for the possible future expansion of their applications. Salient among these challenges are the following, which are also suggested as future research directions, recommendations and projects:

- Training: the requirement to educate a substantial cadre of relevant technical and managerial staff.
- IoT Power: it is imperative to minimize energy consumption in IoT devices and to implement energy harvesting to power them.
- Interoperability Standards: international agreement on interoperability standards.
- Protocol Standardization: international agreement on application layer protocol.
- Pervasiveness: seamless integration of IoT with the pervasive computing community.
- Sensor Technology: expansion of sensor abilities.
- IoT OS: need for an agreed resource allocation strategy.
- IoT Network Architects: establishment of a stable network architecture.
- Fog Computing: development of aspects of fog computing relevant to IoT.
- Hardware Interoperability: standards for interoperability with wireless sensor networks.
- Ethical standards: recognition of human needs for employment, privacy and truthful information.
- Device Hardening: procedures to frustrate hacking, e.g., the need for the widespread use of the blockchain and more secure encryption of IoT data.
- IoT Discovery Protocol: the need for an IoT device discovery protocol. The Shodan search engine can specifically search for IoT devices.
- IoT ID Protocol: need for identity management methodology.
- Beyond Big Data Management: procedures for the management of the huge data quantities already being generated.

Diffusion of any technology has a significant impact on the economy, especially the relationship between investments and "return", which is extremely important. Return on any such technological investment is highly multifaceted and cannot be measured based on just one single aspect such as the economy. For example, the return on investment in IoT in the healthcare sector may not be monetarily justified; however, it has great impact on human society and thus large non-monetary value. Since IoT is still in its infancy, many organizations that are active in IoT in various scenarios still do not have a viable business model for their IoT applications. Though this may appear as an economic challenge, the exponential market penetration of IoT devices shows the probable lucrative trend of this new emerging market. Thus, based on the evidences so far, IoT will very probably enable various new and distinct business models to be developed in the near future. The discussion of such economic aspects of IoT must be seen as an important topic for future research.

Overall, it is evident that a rich and urgent research and development field in IoT has been broadened. It is, therefore, hoped that this review will encourage the development on the themes outlined.

**Author Contributions:** Mahdi H. Miraz: expertise on computer networking, IoT and cloud computing. Maaruf Ali: expertise on mobile communications, computer networking and IoT. Peter S. Excell: expertise on telecommunications engineering aspects, especially concerning wireless links. Richard Picking: expertise on the history and present state of the Internet plus specialist advice on human-machine-interaction.

**Funding:** This research received no external funding.

**Conflicts of Interest:** The authors declare no conflict of interest.